\documentstyle[11pt,aaspp4]{article}

\slugcomment{To appear in ApJ Letters (July 20, 1996)}

\def\be{\begin{equation}}
\def\ee{\end{equation}}
\def\go{\mathrel{\raise.3ex\hbox{$>$}\mkern-14mu
             \lower0.6ex\hbox{$\sim$}}}
\def\lo{\mathrel{\raise.3ex\hbox{$<$}\mkern-14mu
             \lower0.6ex\hbox{$\sim$}}}
\def\Porb{P_{\rm orb}}
\def\vxi{{\vec\xi}}
\def\br{{\bf r}}

\begin{document}

\title{Orbital Decay of the PSR J0045-7319 Binary System: 
Age of Radio Pulsar\\
and Initial Spin of Neutron Star}

\author{Dong Lai}
\affil{Theoretical Astrophysics, 130-33, California Institute of
Technology\\
Pasadena, CA 91125\\
E-mail: dong@tapir.caltech.edu}

\begin{abstract}
Recent timing observations of PSR J0045-7319 reveal that 
the neutron star/B star binary orbit is decaying on a time
scale of $|\Porb/\dot\Porb|=0.5$ Myr, shorter than the characteristic
age ($\tau_c=3$ Myr) of the pulsar (Kaspi et al.~1996a). We study
mechanisms for the orbital decay. The standard weak friction theory
based on static tide requires far too short a viscous time to
explain the observed $\dot\Porb$. We show that dynamical tidal
excitation of g-modes in the B star can be responsible for the orbital
decay. However, to explain the observed short decay timescale,
the B star must have some significant retrograde rotation with
respect to the orbit --- The retrograde rotation brings 
lower-order g-modes, which couple much more strongly to the tidal
potential, into closer ``resonances'' with the orbital motion, thus
significantly enhancing the dynamical tide. A much less likely
possibility is that the g-mode damping time is much shorter than the
ordinary radiative damping time. The observed orbital decay timescale
combined with a generic orbital evolution model based on dynamical
tide can be used as a ``timer'', giving an upper limit of $1.4$ Myr
for the age of the binary system since the neutron star formation.
Thus the characteristic age of the pulsar is not a good age indicator.
Assuming standard magnetic dipole braking for the pulsar and no
significant magnetic field decay on a timescale $\lo 1$ Myr,
the upper limit for the age implies that the initial spin of the
neutron star at birth was close to its current value. 
\end{abstract}


\section{Introduction}

One of the fundamental questions in the study of pulsars
concerns the initial conditions of neutron stars at birth. 
In particular, the initial spin of neutron star is related 
to such issues as angular momentum coupling 
in the progenitor stars, supernova explosion mechanisms
and gravitational wave generation from core collapse and 
nascent neutron stars.
Unfortunately, 
except for the Crab pulsar, for which 
the initial spin can be directly inferred from the measured pulsar
period $P_p$ and its time-derivatives $\dot P_p,\,\ddot P_p$ and the
known age, 
our knowledge about this quantity is rather limited.
Statistical studies of the evolution of pulsar population 
(e.g., Narayan 1987), the energetics of pulsar nebulae
(Helfand \& Becker 1987)
and possible old pulsar/supernova remnant associations
(Kaspi et al.~1996b)
give some indications that neutron stars are formed rotating at 
a moderate spin period $10-100$ ms. However,
all these methods suffer from uncertainties, and  
other independent constraints on the initial spins 
of neutron stars are highly valuable.  

The PSR J0045-7319 binary contains 
a radio pulsar 
($P_p=0.93\,$s) and massive B-star companion in an eccentric, 
$51.17\,$ days orbit (\cite{Kaspi94}). This system 
has recently been shown to exhibit spin-orbit precessions due to the
rapid, misaligned rotation of the B star,
which strongly suggests that the neutron star received a kick 
at birth from asymmetric supernova (\cite{Lai95,Kaspi96a}). 
Recent timing observations also reveal that the orbit is 
decaying on a time scale 
$|\Porb/\dot\Porb|=0.5\,$ Myr, shorter than the pulsar's
characteristic
age $\tau_c=P_p/(2\dot P_p)=3\,$Myr (\cite{Kaspi96a}).
In this paper, we study the physical mechanisms for the rapid 
orbital decay, and demonstrate the potential of using the orbital
decay time to constrain the age and initial spin of the pulsar. 
 
The fiducial numbers we adopt for the current PSR J0045-7319 system
are: pulsar mass $M_p=1.4M_\odot$, companion mass $M_c=8.8M_\odot$, 
radius $R_c=6.4R_\odot$, and orbital period 
$\Porb=51.17$ days, semi-major axis $a=20R_c$, eccentricity $e=0.808$
(\cite{Kaspi96a,Bell95}).

\section{Mechanisms for the Orbital Decay}

Since the mass loss from the B star is completely negligible 
(\cite{Kaspi96a}),
tidal interaction is the most natural cause for the orbital decay
in the PSR J0045-7319 system. 
There are two kinds of tides: static and dynamical.
The static tide corresponds to the global, quadrupolar
distortion of the star. In the presence of finite viscosity, 
a tidal lag angle $\theta_{\rm lag}\propto t_{\rm visc}^{-1}$
develops between the tidal
bulge and the line joining the pulsar and the B star. The orbital
evolution and spin evolution are driven by the viscous tidal
torque $\sim (GM_p^2R_c^5/a^6)\theta_{\rm lag}$. 
However, 
this ``weak friction theory'' based on static tide
cannot explain the orbital decay in the PSR J0045-7319 binary
for two reasons: (i) Using the general tidal 
equations (e.g., Alexander 1973) together with constraints on the 
stellar rotation rate and spin-orbit inclination angle, we have found 
that a viscous time $t_{\rm visc}$ of less than $30$ days is needed in
order to explain the observed $\dot\Porb$; Such a small viscous time
is almost certainly impossible.
(e.g., even if the star were completely convective, the viscous time
would still be longer than one year).
(ii) The same amount of torque that induces the orbital decay also acts
on the B star; Since the stellar spin $S$ is much less than 
the orbital angular momentum $L$, 
the static tide would lead to rapid spin-orbit synchronization
and alignment if it were to explain the observed $\dot\Porb$,
in contradiction with observations.   
In the following, we discuss the physics and outline the main result
of a dynamical tide theory that can explain the observations. 
A more complete analyses will be presented in a future paper. 

At each periastron passage, the pulsar excites non-radial
(mainly $l=2$ quadrupolar) oscillations in the B star,
therefore energy and angular momentum are transferred between the
orbit and the star. Write the Lagrangian displacement in the
rotating frame of the star as 
$\vxi(\br,t)=\sum_\alpha\!a_\alpha(t)\vxi_\alpha(\br)$,
where $\vxi_\alpha(\br)$ is the eigenmode, 
normalized via $\int\!d^3x\,\rho\,\vxi_\alpha^\ast\cdot\vxi_\alpha=1$,
and $\alpha=\{n\lambda m\}$ is the mode index
($n$ is the radial order of the mode, $\lambda$ and $m$ specify
the angular eigenfunction; note that $\lambda=l$ only for nonrotating
star). The mode amplitude $a_\alpha(t)$ is then governed by the
equation
\be
\ddot a_\alpha+\omega_\alpha^2a_\alpha+2\gamma_\alpha\dot a_\alpha
={GM_pW_{lm}Q_\alpha\over D^{l+1}}e^{-im(\Phi-\Omega_st)},
\ee
where $\omega_\alpha$ is the mode angular frequency (in the rotating 
frame), $\gamma_\alpha$ the amplitude 
damping rate, $D(t)$ is the orbital separation,
$\Phi(t)$ the orbital phase, $\Omega_s$ is the 
rotation rate of the companion\footnote{The observation of orbital
precession implies that the stellar spin and the orbital angular
momentum are not aligned. Here we assume they are aligned for
simplicity. Thus $\Omega_s$ should be understood as the component of
the spin rate perpendicular to the orbital plane.}, 
$W_{lm}$ is a constant as defined in Press and Teukolsky (1977), 
and $Q_\alpha$ is the tidal coupling coefficient as given by 
$Q_\alpha=\int\!d^3x\,\rho\,{\vec\xi}_\alpha^\ast\cdot\nabla
(r^lY_{lm})$.

The energy transfer between the star and orbit during each periastron
passage depends on the phases of the oscillation modes, thus in general
varies from one passage to another (e.g., Kochanek 1992; Mardling 1995).
An important measure of the strength of dynamical tide is the energy
$\Delta E_s$ transferred to the star (and the corresponding angular
momentum transfer $\Delta J_s$) during the ``first'' periastron when
there is no oscillation present initially. This can be calculated using
the formalism of Press and Teukolsky (1977), generalized to include
stellar rotation and elliptic orbit. To the leading ($l=2$
quadrupole) order, we have
\be
\Delta E_s={GM_p^2\over R_c}\left({R_c\over
a_p}\right)^6\,T_2(\eta),~~~~
T_2(\eta)\simeq 2\pi^2\sum_\alpha {(Q_\alpha K_{2m,\alpha})^2\over
F_\alpha},
\ee
where $\eta=(M_c/M_t)^{1/2}(a_p/R_c)^{3/2}$ is the ratio of 
the time for periastron passage and the stellar dynamical time, 
$M_t=M_p+M_c$ is the total mass, and
$a_p=a(1-e)$ is the periastron distance.
The factor $F_\alpha=1-(i/\omega_\alpha)\int\!d^3x\,\rho\,
{\vec\xi}_\alpha^\ast\cdot ({\vec\Omega_s}\times{\vec\xi}_\alpha)$
is of order unity, and 
$K_{lm,\alpha}$ depends on the orbital trajectory and the mode frequency:
\be
K_{lm,\alpha}={W_{lm}\over 2\pi}\int_{-\Porb/2}^{\Porb/2}
\!\!\!dt {1\over D^{l+1}}\cos\,[\omega_\alpha
t+m\Phi(t)-m\Omega_s t],
\ee
with the integration centered around periastron.
We can obtain a similar expression for the angular momentum transfer:
\be
\Delta J_s={GM_p^2\over R_c}\left({R_c\over a_p}\right)^6
\!\left({R_c^3\over GM_c}\right)^{1/2}\!\!S_2(\eta),~~~~
S_2(\eta)\simeq 2\pi^2\sum_\alpha\left({-m\over\omega_\alpha}\right)
{(Q_\alpha K_{2m,\alpha})^2\over F_\alpha}.
\ee
Clearly, the most strongly excited modes which contribute 
most to the energy transfer are those 
(i) propagating in the same direction as the orbit 
(the $m=-2$ modes); 
(ii) having frequencies in the inertial frame
$(\omega_\alpha-m\Omega_s)\sim (-m\dot\Phi_p)$,
where $\dot\Phi_p=\Omega_{\rm orb}
(1+e)^{1/2}/(1-e)^{3/2}=2\pi/(3.2\,{\rm days})$ is the orbital
frequency at periastron (``resonance'' condition); and 
(iii) having relatively large $Q_\alpha$.
For nonrotating star, these modes are g-modes
of radial order $n=5-9$ (i.e., $g_5-g_9$). 

The rapid rotation of the B star changes its g-modes significantly. 
Since the spin frequency is comparable to the relevant g-mode
frequencies, a perturbation treatment of the rotation effect is not
valid. Neglecting the centrifugal force, the g-modes of rotating
star can be calculated using the ``traditional approximation'' 
(e.g., Unno et al.~1989), where the Coriolis force associated with
the horizontal component of the spin angular velocity is ignored
(valid because the horizontal displacement of the g-mode is larger
than the radial displacement). We correct the g-mode
frequencies of a $\Gamma=4/3,\,\Gamma_1=5/3$ nonrotating
polytrope\footnote{This is appropriate for a massive star, although 
the g-modes depend somewhat on the size of the convective core,
and hence on the stellar age.
We have tested that $|\Delta E_s|$ is rather insensitive to the
core size.}
using the numerical result of Berthomieu et al.~(1978) to
obtain the mode frequencies of the rotating star. 
Since $|\omega_\alpha/\Omega_s|$ is greater than unity (although not 
much greater) for the modes that are strongly excited, 
the rotational corrections to the mode frequencies and wavefunctions
are not large (less than $40\%$ in $\omega_\alpha$). 
We therefore adopt $Q_\alpha$ to be approximately the same as 
that of the non-rotating star; its validity is also indicated 
by an asymptotic analysis (Rocca 1987).

Figure 1 shows the function $T_2(\eta)$ for different values of 
$\Omega_s$. We see that {\it rotation can significantly change
the strength of dynamical tide}. In particular, at 
$\eta=7$ (appropriate for the current PSR J0045-7319 system), 
retrograde rotation ($\Omega_s<0$) can increase $\Delta E_s$ from the
nonrotating value by
more than two orders of magnitude. Physically, this dramatic effect
of rotation on the tidal strength comes about because 
(i) the retrograde rotation brings lower-order g-modes, which 
couple to the tidal potential more strongly, into closer ``resonances'' 
with the orbital motion; (ii) The tidal coupling coefficient
$Q_\alpha$ depends very strongly on the order of the mode. 
For example, for $\hat\Omega_s\equiv\Omega_s(GM_c/R_c^3)^{-1/2}=-0.4$,
the most strongly excited modes are $g_4-g_5$, while
$g_1-g_3$ are also much more strongly excited compared to the 
nonrotating case. 
The function $S_2(\eta)$ has similar behavior as $T_2(\eta)$, 
and we find that typically $S_2\simeq 2T_2$ to within $10\%$
(for $|\hat\Omega_s|<0.5$). 

The radiative damping times of f-mode and low-order g-modes 
of a massive main sequence star range from 10's to 100's of years
(Saio \& Cox 1980; Dziembowski et al.~1993), 
much longer than $\Porb$. 
Figure 2 shows an example of the dynamical tidal evolution 
obtained by integrating Eq.~(1) and the orbital equations. 
The average mode energy settles into a constant value 
$E_s\simeq\Delta E_s$ (a ``fixed point'' of the dynamical system) 
after a few damping times, and the orbital energy decreases secularly
according to
\be
\dot E_{\rm orb}\simeq -2\gamma |\Delta E_{\rm orb}|
=-2\gamma (\Delta E_s+\Omega_s\Delta J_s),
\ee
where $\gamma_\alpha=\gamma=1/t_{\rm damp}$
(i.e., we assume that the damping times are the same
for different modes; this is approximately valid since only a 
small number of modes are strongly excited).
With $S_2\simeq 2T_2$, we have $\Delta J_s\simeq
2(R_c^3/GM_c)^{1/2}\Delta E_s$, and 
the secular orbital decay rate is then given by 
\be
{\dot\Porb\over\Porb}\simeq -{1+2\hat\Omega_s\over 0.24\,{\rm Myr}}
\left({100\Porb\over t_{\rm damp}}\right)
\left({T_2\over 10^{-2}}\right).
\ee
For nonrotating star, we find $T_2\simeq 2.6\times 10^{-4}$, thus 
a very short damping time $t_{\rm damp}\simeq 5.5\Porb$ is needed to
explain the observed $\dot\Porb$. However, when the star has significant 
retrograde rotation, the observed orbital decay rate can be explained
with ordinary radiative damping.

If the mode damping time is shorter than the orbital period, then
the secular orbital decay rate should be given by 
$\dot\Porb\sim -|\Delta E_{\rm orb}|/|E_{\rm orb}|$.
This would require significant enhancement to the mode dissipation 
compared to ordinary radiative damping.
Nonlinear mode damping (Kumar \& Goodman 1996) is unimportant
since the mode energy $E_s/(GM_c^2/R_c)\sim
10^{-7}(t_{\rm damp}/100\Porb)$, 
implied from the observed orbital decay rate, is much
smaller than the threshold $E_{s,th}/(GM_c^2/R_c)\sim 
(\omega_\alpha t_{th})^{-0.4}\sim 10^{-3}$ needed for the three-mode
parametric resonant coupling (where $t_{th}\sim 10^4$ yr is the thermal 
time of the star, and we have approximated the damping time of 
the $n\sim l>>1$ mode by $t_{th}/l^8$).
We consider this possibility rather unlikely. 

\section{Orbital Decay Model and Constraints on the Pulsar
Age and Initial Spin}

We now consider the long-term evolution of the binary system.
Equations (2) and (5) provide a scaling relation for 
the orbital decay rate. Let $T_2(\eta)\propto\eta^{-4\nu}$
(with $\nu\simeq 1,\,0.5,\,0.2$ for $\hat\Omega_s=0,\,-0.2,\,-0.4$
respectively), we obtain
\be
\dot\Porb=-A\Porb^{-7/3-4\nu}(1-e)^{-6(1+\nu)},
\ee
where the constant $A$ can be fixed by the observed current
$\dot\Porb$ value. 
Using $\dot L=-\dot J_s\simeq
-2(R_c^3/GM_c)^{1/2}\dot E_s$, we have
\be
{\dot L\over L}={1\over\beta}{P_{{\rm orb},0}(1-e_0^2)^{1/2}\over\Porb
(1-e^2)^{1/2}}{2\dot\Porb\over 3\Porb},
\ee
where the subscript ``0'' indicates current values, and the parameter
$\beta\simeq 50(1+2\hat\Omega_s)$ ranges from $~10$ to $50$
for $-0.4<\hat\Omega_s<0$. Note that the 
long-term orbital angular momentum change\footnote{The orbital angular
momentum is transferred into the spin, whose evolution 
can be easily evaluated: Since $\dot\Omega_s/\Omega_s
\simeq (0.04/\hat\Omega_s)(\dot \Porb/\Porb)$ (with the moment of 
inertia $0.1M_cR_c^2$),
the timescale for changing the stellar spin (both its magnitude and 
direction) is longer than the orbital decay timescale. 
This feature is qualitatively different from static tide.}
is much less than the change in the orbital energy since 
$|\Delta J_s/L|\simeq 
|\Delta E_s/E_{\rm orb}|\Omega_{\rm orb}(GM_c/R_c^3)^{-1/2}
(1-e^2)^{-1/2}<<|\Delta E_s/E_{\rm orb}|$.
The evolution of eccentricity is given by 
\be
\dot e={1-e^2\over e}\left({\dot\Porb\over 3\Porb}-
{\dot L\over L}\right).
\ee
For the current PSR J0045-7319 binary, we find
$\dot e\simeq 10^{-14}$ s$^{-1}$, consistent with the 
observational upper limit for $\dot e$ (Kaspi et al.~1996a).

Equations (7)-(9) can be integrated backward to give 
$\Porb$ and $e$ at earlier times\footnote{We assume there is
no major structure change in the B star during the orbital evolution.
This is most likely to be valid because the Kelvin-Holmholtz timescale
($\sim 10^4$ years) is much shorter than the orbital evolution time.}.
The parameters $\nu$ and $\beta$ 
have been chosen to reflect their allowed ranges.
The results are shown in Figure 3. We see that $e\rightarrow 1$ 
and $\Porb\rightarrow\infty$ as $(-t)$ approaches a fixed value 
$\lo 1.4$ Myr, independent of the parameters $\nu$ and $\beta$. 
This feature can be understood as follows: Since the 
angular momentum transfer in a periastron passage is
relatively small compared to the energy transfer, 
the periastron distance $a_p$ 
and hence $\Delta E_s$ remain approximately constant 
during the evolution. Thus $|\dot\Porb/\Porb|\propto
\Delta E_s/|E_{orb}|$ is larger at earlier times. 

The observed $\dot\Porb$ and our generic orbital evolution model 
therefore set an upper limit $\simeq 1.4$ Myr to 
the age of the binary system since the last supernova
\footnote{In the unlikely case of $t_{\rm damp}\lo\Porb$,
the scaling relation for $\dot\Porb$ is modified from Eq.~(7)
to $\dot\Porb\propto -\Porb^{-10/3-4\nu}(1-e)^{-6(1+\nu)}$,
and an absolute upper limit to the age cannot be obtained
(see the dotted line in Fig.~3). However, the age can still be
constrained since the probability of forming a highly eccentric,
bound orbit after supernova is small.}.
This is significantly smaller than the 
characteristic age $\tau_c$ of the pulsar, implying that the latter 
is not a good age indicator. In principle, the
discrepancy between the age and $\tau_c$
may be explained in several ways: (i) The pulsar has been 
slowed down by an earlier phase of mass accretion; 
(ii) The pulsar braking index is much larger than $3$
(the canonical dipole value); (iii) The pulsar 
magnetic field decays on a timescale $\lo 1$ Myr; 
(iv) The initial spin period of the pulsar is not much shorter 
than the current value. The extremely small mass loss from the 
B star (Kaspi et al.~1996a) argues strongly against (i). 
The observed braking indices measured for three of the youngest 
pulsars range from $2.0$ to $2.8$, and in view of the statistical
evidence and physical argument against rapid field decay, 
we consider (ii) and (iii) 
to be rather unlikely.  Using the standard magnetic dipole braking 
for the pulsar spin-down, the constraint on the pulsar age 
implies that the initial spin period of the pulsar is longer than 
$0.7$ second, very close to its current value. 
Thus PSR J0045-7319 was formed rotating very slowly,
and its progenitor 
must also have rather slow rotation. We note that such a small initial
spin rate of the pulsar may result from the random torque on
the nascent neutron star due to asymmetry in the supernova
(e.g., Burrows et al.~1995).

\section{Discussion}

Our analyses of PSR J0045-7319 binary orbital decay show
that the B star companion is most likely to have a retrograde 
rotation with respect to the orbit. 
Since the spin of the B star is expected to have been aligned with 
the orbital angular momentum before the supernova,
the current misaligned configuration could have come about 
only if the neutron star received a kick at birth due to
asymmetry in the supernova. The kick velocity must 
(i) have had a component out of the original 
orbital plane in order to misalign ${\bf L}$ and ${\bf S}$;
(ii) have had a significant component in the direction opposite to
the orbital velocity of the progenitor in order to reverse
${\bf L}$.
The current timing data yield two degenerate solutions 
for the range of the spin-orbit inclination angle, allowing for
both prograde and retrograde rotation (Kaspi et al.~1996a).  
Long-term (i.e., some fraction of the precession period $\sim 500$
years) timing observation 
should distinguish these two possibilities. Dedicated
optical observation of the companion may also give
useful constraints on the excited modes (Kumar et al.~1995).

Our analyses have also demonstrated that the orbital decay can be
used to put concrete constraints on the pulsar age and initial spin. 
The tide-induced orbital decay of the PSR B1259-63/Be star binary 
(the only other known radio pulsar/main sequence star binary)
is too slow to be observable owing to its larger orbital separation
at periastron. Finding more systems similar to PSR J0045-7319 
will allow for determination of systematic constraints on the physical
conditions of neutron stars at birth and supernova characteristics.

\acknowledgments

I thank Vicky Kaspi for sharing the observational data with me 
and helpful discussion and comments. I also thank Lars
Bildsten, Peter Goldreich, Norm Murray and Yanqin Wu for useful
discussion. This research is supported by the Richard C. Tolman
Fellowship at Caltech, NASA Grant NAG 5-2756, and NSF Grant
AST-9417371. 


\clearpage

\figcaption
{The function $T_2(\eta)$ (defined in Eq.~[2]) for different rotation
rates: $\hat\Omega_s=0$ (solid lines), $0.2$ (dotted lines),
$-0.2$ (short-dashed lines), and $-0.4$ (long-dashed lines).
Negative $\hat\Omega_s$ corresponds to retrograde rotation. 
The heavy lines are for elliptic orbit with $e=0.808$
(the current value for the PSR J0045-7319 system), 
and the light lines for parabolic orbit.
\label{fig1}}

\figcaption
{Evolution of $\delta E_{orb}\equiv E_{orb}-E_{orb,0}$ and 
$E_s$ (mode energy in the rotating frame) due to 
dynamical tide (with $\hat\Omega_s=-0.2$). 
Only the $g_3-g_8$ modes are included in the
calculation, and relatively large damping rates are chosen
for clearer illustration:
$\gamma=0.1/\Porb$ (solid curves) and $\gamma=0.05/\Porb$
(dotted curves). 
\label{fig2}}

\figcaption
{The long-term (backward) evolution of the binary period and
eccentricity ($t=-3$ Myr
corresponds to the pulsar's characteristic age). The solid lines
are for $\beta=10$, with $\nu=0.2,~0.5,~1$ (left to right),
the short-dashed lines are for $(\nu,\beta)=(0.5,20)$,
the long-dashed lines for $(\nu,\beta)=(0.5,50)$. The 
dotted curves are for a model assuming $t_{\rm damp}\lo\Porb$
with $(\nu,\beta)=(1,50)$ (see footnote 5).
\label{fig3}}

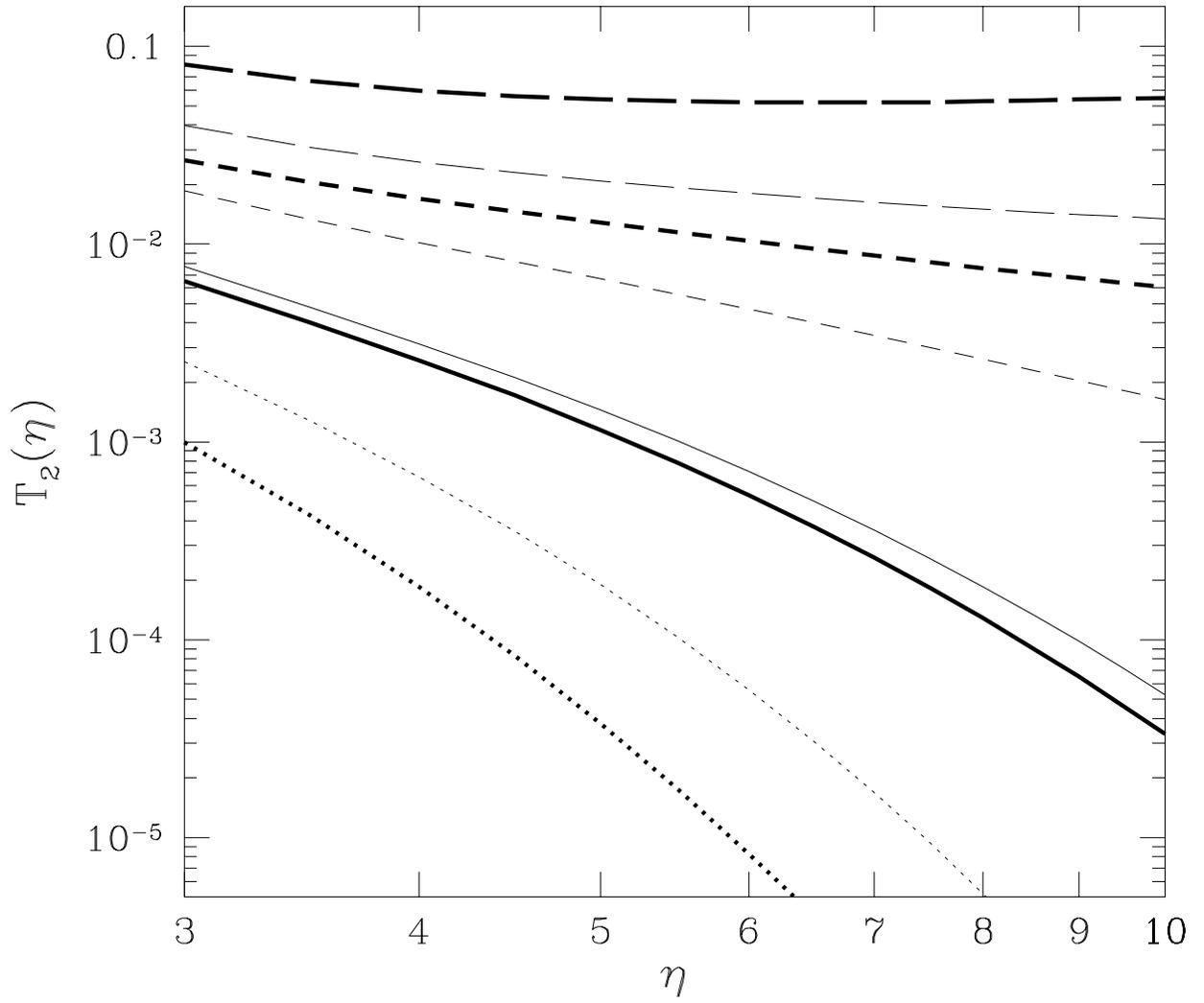
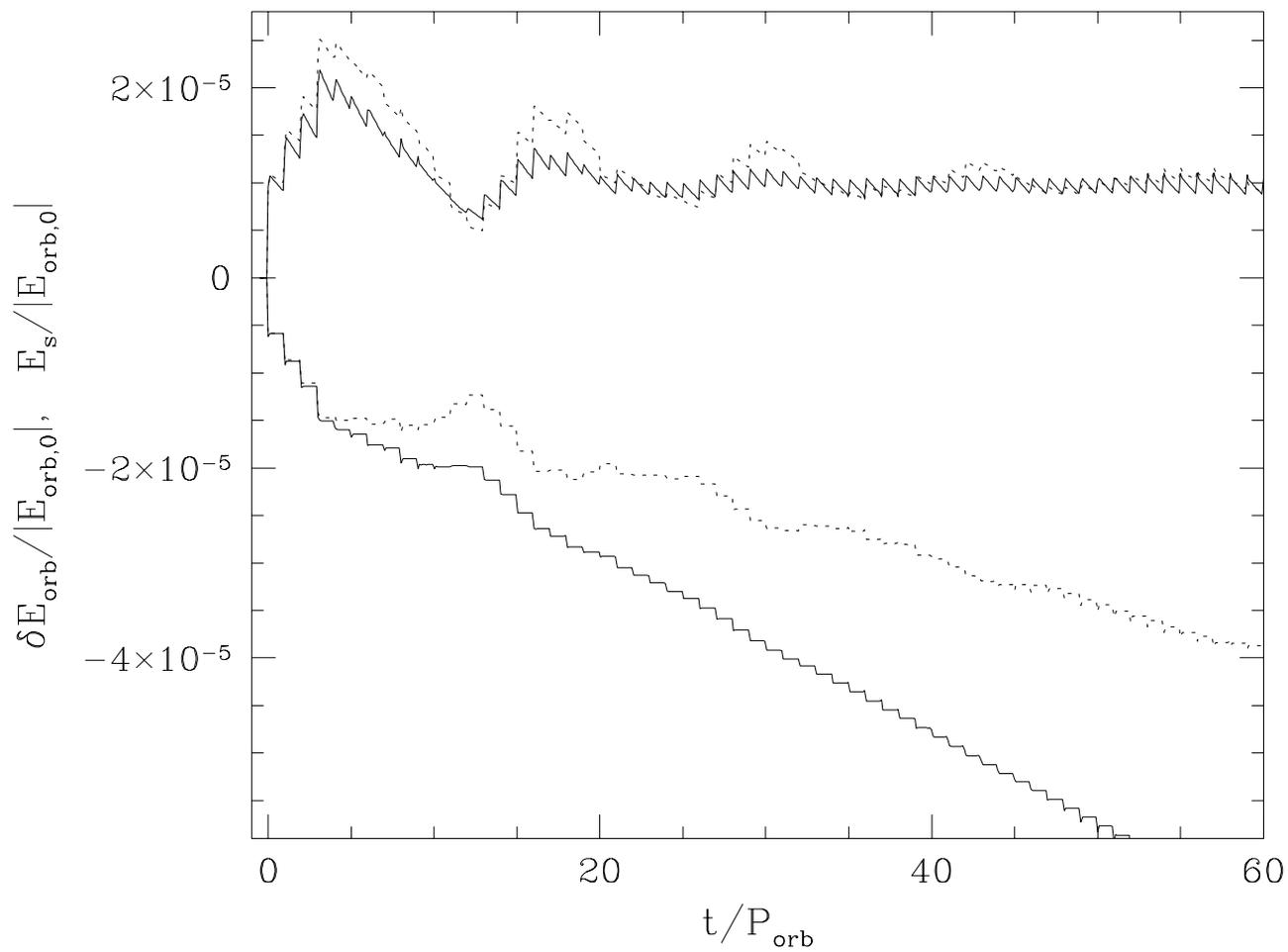
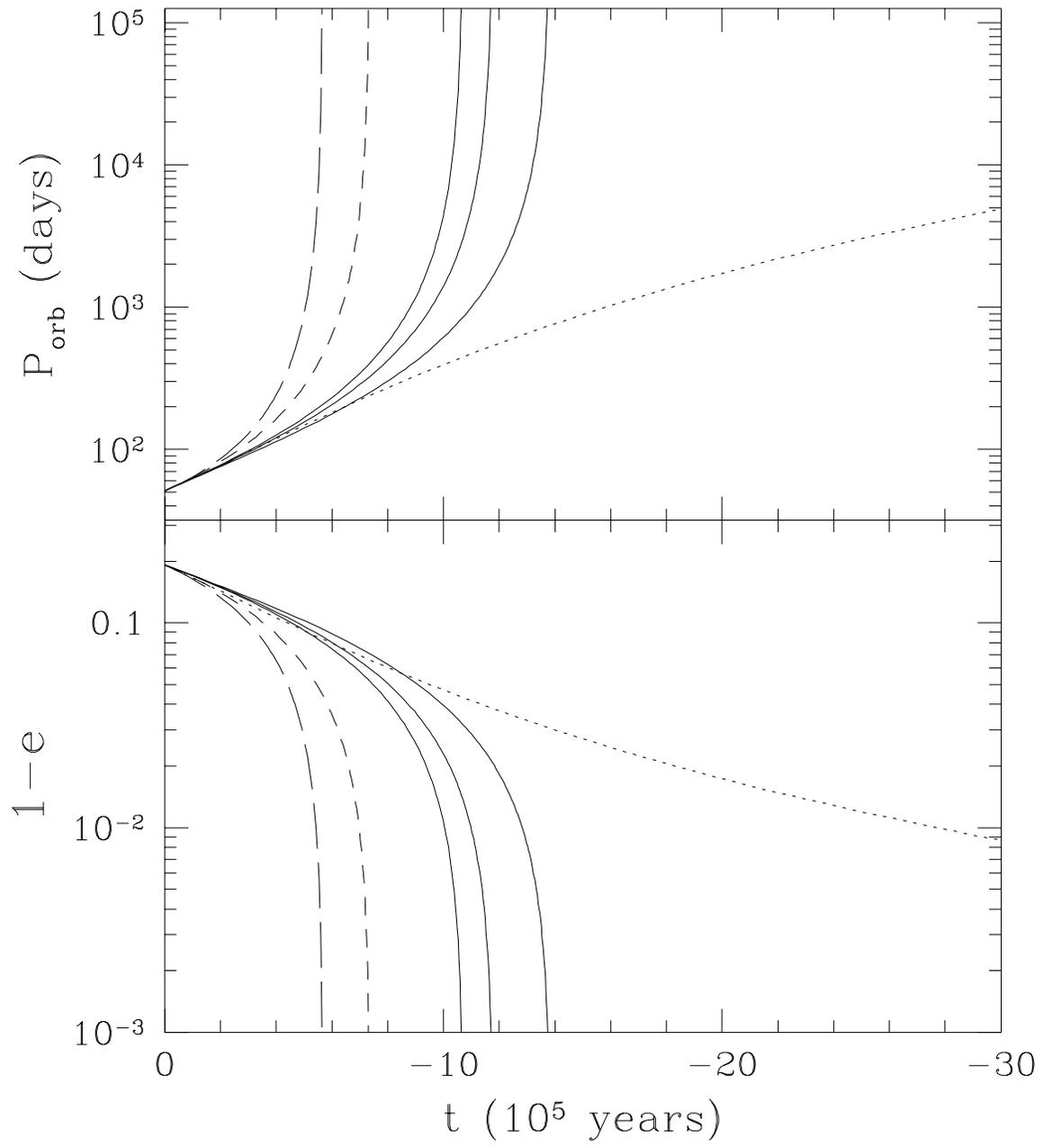
\end{document}